\documentclass[12pt]{article}
\usepackage{graphicx}
\usepackage{epsfig}
\begin {document}
\begin{center}
\bf
{\Large Odderon Effects in pp Collisions:\\ Predictions for LHC Energies}

\vspace{.2cm}

C. Merino$^*$, M. M. Ryzhinskiy$^{**}$, and Yu. M. Shabelski$^{**}$ \\

\vspace{.5cm}
$^*$ Departamento de F\'\i sica de Part\'\i culas, Facultade de F\'\i sica, \\ 
and Instituto Galego de F\'\i sica de Altas Enerx\'\i as (IGFAE), \\ 
Universidade de Santiago de Compostela, \\
Santiago de Compostela 15782, \\
Galiza, Spain \\
E-mail: merino@fpaxp1.usc.es

 

\vspace{.2cm}

$^{**}$ Petersburg Nuclear Physics Institute, \\
Gatchina, St.Petersburg 188350, Russia \\
e-mail: mryzhinskiy@phmf.spbstu.ru \\
e-mail: shabelsk@thd.pnpi.spb.ru
\vskip 0.9 truecm

\centerline{\it Lecture given by Yu.M. Shabelski at the XLIII PNPI Winter School on Physics}

\centerline{\it St.Petersburg, Russia}
\vskip 0.3 truecm

\centerline{\it February 2009}
\vskip 0.9 truecm

A b s t r a c t
\end{center}

We consider the possible contribution of Odderon (Reggeon with $\alpha_{Odd}(0) \sim 1$ 
and negative signature) exchange to the differences in the total cross sections of 
particle and antiparticle, to the ratios of real/imaginary parts of the elastic $pp$ 
amplitude, and to the differences in the inclusive spectra of particle and antiparticle in 
the central region. The experimental differences in total cross sections of particle and
antiparticle are compatible with the existence of the Odderon component but such a large Odderon
contribution seems to be inconsistent with the values of Re/Im ratios. In the case of inclusive 
particle and antiparticle production the current energies and/or accuracy of 
the experimental data don't allow a clear conclusion. It is expected that the LHC will finally
solve the question of the Odderon existence.

\newpage

\section{Introduction}

The Odderon is a singularity in the complex $J$-plane with intercept 
$\alpha_{Odd} \sim 1$, negative $C$-parity, and negative signature. Thus its 
zero flavour-number exchange 
contribution to particle-particle and to antiparticle-particle interactions, 
e.g., to $pp$ and $\bar{p}p$ total cross sections, has opposite signs. 
In QCD the Odderon singularity is connected~\cite{BLV} 
to the colour-singlet exchange of three reggeized gluons in $t$-channel.
The theoretical and experimental status of Odderon has been recently discussed 
in refs.~\cite{Nic,Ewe}. The possibility to detect Odderon
effects has also been investigated in other domains as
$\pi p \to \rho N$ reaction~\cite{Cont},
charm photoproduction~\cite{BMR}, and diffractive production of pion pairs~\cite{Hag,Ginzb}.

The difference in the total cross sections of antiparticles and particles 
interactions with nucleon targets are numerically small and decrease rather fast 
with initial energy, so the Odderon coupling should be very small with respect to
the Pomeron coupling. However, several experimental facts favouring the presence of
the Odderon contribution exist. One of them is the difference in the 
$d\sigma/dt$ behaviour of elastic $pp$ and $\bar{p}p$ scattering at $\sqrt{s} =$
52.8 GeV and $\vert t \vert = 1 - 1.5$ GeV$^2$ presented in references~\cite{Nic,Bre}.
The behaviour of $pp$ and $\bar{p}p$ elastic scattering
at ISR and SPS energies was analyzed in~\cite{JSS}. 
Also the differences in the yields of baryons and antibaryons produced 
in the central (midrapidity) region and in the forward hemisphere in meson-nucleon 
and in meson-nucleus collisions, and in the midrapidity region of high energy $pp$ 
interactions \cite{ACKS,BS,AMS,Olga,SJ3,AMS1,MRS}, can also be significant in this respect.
The question of whether the Odderon exchange is needed for explaining 
these experimental facts, or they can be described by the usual exchange of a reggeized quark-antiquark pair with 
$\alpha_{\omega}(t) = \alpha_{\omega}(0) + \alpha_{\omega}'t$ ($\omega$-Reggeon exchange) is a fundamental one.

The detailed description of all available data on hadron-nucleon elastic scattering with accounting for Regge cuts
results in $\alpha_{\omega}(0) = 0.43$, $\alpha_{\omega}' = 1$ GeV$^{-2}$ \cite{VLLTM},
and the simplest power fit
\begin{equation}
\Delta \sigma_{hp}\ = \sigma^{tot}_{\bar{h}p} - \sigma^{tot}_{hp} =
\sigma_R\cdot (s/s_0)^{\alpha_R(0) - 1}
\end{equation}
for experimental points of $\bar{p}p$ and $pp$ scattering starting from $\sqrt{s} = 5$ GeV
gives the value $\alpha_R = 0.424 \pm 0.015$ \cite{ShSh}.
The accounting for Regge cut contributions of the type $RP$, $RPP$, $RP...P$, and $Rf$, $RfP$, $RfP...P$ slightly 
decrease (from  $\alpha_R = 0.424 \pm 0.015$ to $\alpha_R = 0.364 \pm 0.015$)~\cite{ShSh}
the effective value of $\alpha_R$. Thus any process with exchange 
of a negative signature object with effective intercept $\alpha_{eff} > 0.7$ could be considered as an Odderon contribution,
while if $\alpha_{eff} \leq 0.5$
one could say that there is no room for the Odderon contribution.

In this paper we carry out this anlysis for the case of high energy $hp$ collisions.
In Section~2 we study the Regge pole contributions from the data on $\bar{p}p$, $pp$, $\pi^{\pm}p$, and $K^{\pm}p$ 
total cross sections. In Section 3 we consider the possible 
Odderon effect on the ratios of real/imaginary parts of the elastic $pp$ amplitude. 
In Section 4 we compare the experimental ratios of $\bar{h}$ to $h$ inclusive production
in the midrapidity (central) region for $pp$ collisions to predictions from double-Reggeon diagrams.
In Sections 5, 6, and 7 we compare these experimental data to the theoretical predictions of the Quark--Gluon String Model (QGSM).
Our predictions for the Odderon effects at LHC energies are presented in Section 8.

\section{Regge-pole analysis of total $hp$ and $\bar{h}p$ cross sections}

Let us start from the analysis of high energy elastic particle and antiparticle scattering 
on the proton target. Here the simplest contribution is the one Regge-pole $R$ exchange
corresponding to the scattering amplitude
\begin{equation}
A(s,t) = g_1(t)\cdot g_2(t)\cdot \left(\frac{s}{s_0}\right)^{\alpha_R(t) - 1}\cdot \eta(\Theta) \;,
\end{equation}
where $g_1(t)$ and $g_2(t)$ are the couplings of a Reggeon to the beam and target hadrons, $\alpha_R(t)$ is the $R$-Reggeon trajectory, and $\eta(\Theta)$ is the signature factor 
which determines the complex structure of the scattering amplitude ($\Theta$ equal 
to +1 and to $-1$ for reggeon with positive and negative signature, respectively):
\begin{equation}
\eta(\Theta) = \left\{ \begin{array}{ll} i - \tan^{-1}(\frac{\pi \alpha_R}2) & \Theta = +1 \\
                 i + \tan({\frac{\pi \alpha_R}2}) & \Theta = -1 \;, \end{array} \right.
\end{equation}
so the amplitude $A(s,t=0)$ becomes purely imaginary for positive 
signature and purely real for negative signature when $\alpha_R \to 1$. 

The contribution of the Reggeon exchange with positive signature is the same for a particles and its antiparticle,
but in the case of negative signature the two contributions have
opposite signs, as it is shown in Fig.~1.
\begin{figure}[htb]
\centering
\vskip -.2cm
\includegraphics[width=.7\hsize]{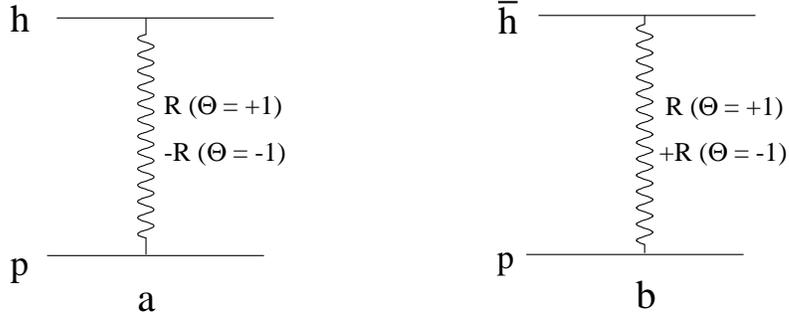}
\vskip -.5cm
\caption{\footnotesize
Diagram corresponding to the Reggeon-pole exchange in particle $h$ (a) (its antiparticle $\bar{h}$
(b)) interactions with a proton target. The positive signature ($\Theta = +1$) 
exchange contributions are the same, while the negative signature ($\Theta = -1$) 
exchange contributions have opposite signs.}
\end{figure}

The difference in the total cross section of high energy particle and antiparticle 
scattering on the proton target is
\begin{equation}
\Delta \sigma^{tot}_{hp} = \sum_{R(\Theta=-1)} 2\cdot Im\,A(s,t=0) =
\sum_{R(\Theta=-1)} 2\cdot g_1(0)\cdot g_2(0)\cdot  \left(\frac{s}{s_0}\right)^{\alpha_R(0) - 1}
\cdot Im\,\eta(\Theta=-1) \;.
\end{equation}

The experimental data for the differences of $\bar{p}p$ and $pp$ total cross sections 
are presented in Fig.~2. Here we use the data compiled in ref.~\cite{CERN} by 
presenting at every energy the experimental points for $pp$ and $\bar{p}p$ by the same experimental
group and with the smallest error bars. At ISR energies (last three points 
in Fig.~2) we present the data in ref.~\cite{Car} as published in their most recent 
version.
\begin{figure}[htb]
\centering
\vskip .2cm
\includegraphics[width=.9\hsize]{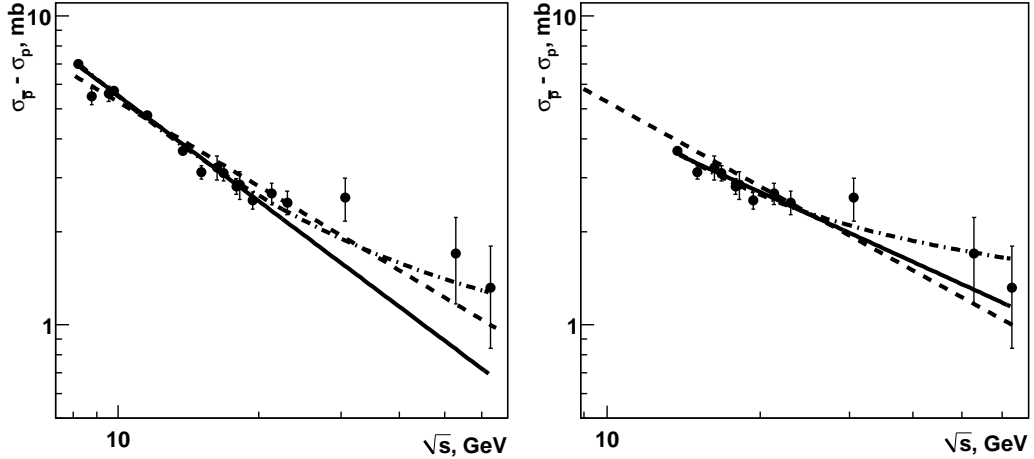}
\vskip -.5cm
\caption{\footnotesize 
Experimental differences of $\bar{p}p$ and $pp$ total cross sections at $\sqrt{s} > 8$ GeV
(left panel) and at $\sqrt{s} > 13$ GeV (right panel) together with their fit by Eq.~(1)
(solid curves), fit of \cite{DL} Eq.~(5) (dashed curves) and fit by Eq.~(6) (dash-dotted curves).}
\end{figure}

In the left panel of Fig.~2 our fit to the experimental data with Eq.~(1) starting 
from $\sqrt{s} > 8$ GeV is presented (solid line). For this fit we obtain the value of 
${\alpha_R} = 0.43 \pm 0.017$ with $\chi^2 =$ 33.3/15 ndf. This result is in good 
agreement with \cite{ShSh}, where the experimental points at energies $\sqrt{s} > 5$~GeV
were included in the fit, and it only slightly differs from the general fit of 
all $hp$ total cross sections \cite{DL} which results in
\begin{equation}
\Delta \sigma_{pp} = 42.31 \cdot s^{-0.4525} {\rm (mb)} \;,
\end{equation}
and it starts from $\sqrt{s} > 10$ GeV. This last fit is also shown in Fig.~2 by a 
dashed line. The values of the parameters for the two fits, together with the $\chi^2$ 
values, are presented in Table~1. It is needed to note that the fit in \cite{DL} was 
aimed at the total $\bar{p}p$ and $pp$ cross sections and not specifically at their 
differences, so the not that good values of $\chi^2$ for this fit are not very 
significant.

As one can see in Table~1, the Eq.~(1) fit can only describe the experimental difference in
the total $\bar{p}p$ and $pp$ cross 
sections when starting from high enough energies. When starting at lower energies other Regge poles,
as well as other contributions, can contribute, but their contribution becomes negligible at higher 
energies. Thus the values of the parameters in Eq.~(1) can be different in different energy regions.
To check the stability of the parameter values, we present in the right panel of Fig.~2 the same experimental data
as in the left panel, but at $\sqrt{s} > 13$ GeV. Here we obtain ${\alpha_R} = 0.62 \pm 0.05$ with $\chi^2 =$ 8.3/10 n.d.f., i.e. now the
description of the data is better, with the value of ${\alpha_R}$ significantly increasing. This indicates that 
it is reasonable to account for two contributions to $\Delta \sigma_{pp}$, the first one corresponding to the well-known $\omega$-reggeon
and the second one corresponding to a possible Odderon exchange:
\begin{equation}
\Delta \sigma_{hp}\ = \sigma_{\omega}\cdot (s/s_0)^{\alpha_{\omega}(0) - 1} +
\sigma_{Odd}\cdot (s/s_0)^{\alpha_{Odd}(0) - 1} \:.
\end{equation}
\begin{center}
\begin{tabular}{|c||r|r|r|} \hline
Parameterization        & $\sigma_R (mb)$ & $\alpha_R(0)$\hspace{0.75cm} & \hskip0.5cm $\chi^2/ndf$ \\  \hline
$p^{\pm}p, \sqrt{s} > 8$ GeV (Eq.~(1))& $75.4 \pm 6.1$  & $0.43 \pm 0.017$  &  35.1/15  \\
$p^{\pm}p, \sqrt{s} > 13$ GeV (Eq.~(1))& $25.5 \pm 7.1$ & $0.625 \pm 0.05$   &  8.8/10   \\
$p^{\pm}p, \sqrt{s} > 8$ GeV (Eq.~(5))& 42.31 (fixed)  & 0.5475 (fixed)     &  92.3/17  \\
$p^{\pm}p, \sqrt{s} > 13$ GeV (Eq.~(5))& 42.31 (fixed) & 0.5475 (fixed)     &  34.5/12   \\
$\pi^{\pm}p, \sqrt{s} > 8$ GeV (Eq.~(1))& $9.51 \pm 1.89$  & $0.51 \pm 0.04$  &  17.2/20  \\
$\pi^{\pm}p, \sqrt{s} > 8$ GeV (Eq.~(5))& 8.46 (fixed)  & 0.5475 (fixed)     &  26.3/22  \\
$K^{\pm}p, \sqrt{s} > 8$ GeV (Eq.~(1))& $28.0 \pm 3.7$  & $0.45 \pm 0.03$  &  15.4/18  \\
$K^{\pm}p, \sqrt{s} > 8$ GeV (Eq.~(5))& 8.46 (fixed)  & 0.5475 (fixed)     & 50.1/20  \\
\hline
\end{tabular}
\end{center}
\noindent
Table 1: {\footnotesize The Regge-pole fits of the differences in $\bar{h}p$ and $hp$ total cross sections
by using Eq.~(1) and Eq.~(5).}

The accuracy of the available experimental points is not good enough for the determination 
of the values of the four parameters in Eq.~(6), so by sticking to the idea of existence 
of the Odderon, we have fixed the value of $\alpha_{Odd}(0)$ close to one
(we take $\alpha_{Odd}(0) = 0.9$\footnote{The value of $\alpha_{Odd}(0)$ can {\it a priori}
be larger than one.}),
thus obtaining the fit shown by a dash-dotted curve both in the 
left panel and in the right panel of Fig.~2 with the values of the parameters presented 
in Table 2.

\begin{center}
\vskip 5pt
\begin{tabular}{|c||r|r|r|r|r|} \hline
Energy & $\sigma_{\omega} (mb)$ & $\alpha_{\omega}(0)$\hspace{0.75cm} & $\sigma_{Odd} (mb)$  
& $\alpha_{Odd}(0)$ & $\chi^2$/ndf  \\   \hline
$p^{\pm}p, \sqrt{s} > 8$ GeV & $165 \pm 37$  & $0.19 \pm 0.06$  & $2.65 \pm 0.45$ 
& 0.9 (fixed) & 26.7/14  \\
$p^{\pm}p, \sqrt{s} > 13$ GeV & $450 \pm 119$ & $-0.09 \pm 0.03$ & $3.61 \pm 0.09$ 
&  0.9 (fixed) & 5.8/9  \\
$p^{\pm}p, \sqrt{s} > 8$ GeV & $172 \pm 7$  & $0.15 \pm 0.09$  & $5.16 \pm 0.32$ 
& 0.8 (fixed) & 27.6/14  \\
$p^{\pm}p, \sqrt{s} > 13$ GeV & $450 \pm 164$ & $-0.16 \pm 0.05$ & $7.38 \pm 0.66$ 
&  0.8 (fixed) & 6.1/9  \\
$\pi^{\pm}p, \sqrt{s} > 8$ GeV & $20.8 \pm 42.6$  & $0.25 \pm 0.60$  & $0.52 \pm 0.69$ 
&  0.9 (fixed) & 16.7/19  \\
$K^{\pm}p, \sqrt{s} > 8$ GeV & $23. \pm 11.2$ & $0.52 \pm 0.86$  & $-0.55 \pm 1.72$ 
&  0.9 (fixed) & 15.3/17  \\
\hline
\end{tabular}
\end{center}
\noindent
Table 2: {\footnotesize The double Regge-pole fit to the differences in $\bar{h}p$
and $hp$ total cross sections using Eq.~(6).}

From the results of this fit for $\sqrt{s} > 8$ GeV one can see that an Odderon
contribution with $\alpha_{Odd}(0) \sim 0.9$ is in agreement with the experimental
data, the values of $\chi^2$/ndf for parametrization by Eq.~(6) being smaller that those
in the case of Eq.~(1). The contributions of Odderon and $\omega$-reggeon to the 
differences in $\bar{p}p$ and $pp$ total cross sections would be approximately equal 
at $\sqrt{s} \sim 25$$-$$30$ GeV. The fit with Eq.~(6) at $\sqrt{s} > 13$ GeV
 qualitatively results in the same curve as the fit at $\sqrt{s} > 8$, but now the 
errors in the values of the parameters are very large.

Such large value of $\alpha_{Odd}(0)$ ($\alpha_{Odd}(0) \sim 0.9$) with a rather
large Odderon coupling should necessarily reflect in a large value of the ratio
\begin{equation}
\rho = \frac{Re A(s,t=0)}{Im A(s,t=0)}\; ,
\end{equation}
but this could be in disagreement with the existing experimental data, as it will be
discussed in the next section. In any case, this problem fades away when 
considering smaller values of $\alpha_{Odd}(0)$. For this reason in Table 2 we present 
our fits for the differences in $\bar{h}p$ and $hp$ total cross sections by using 
Eq.~(6) with a fixed value $\alpha_{Odd}(0) = 0.8$. The new curves are very close 
to those of the $\alpha_{Odd}(0) = 0.9$ fit, but now the values of $\chi^2$/ndf are 
slightly increased.

In the left panel of Fig.~3 the experimental data for the differences of $\pi^- p$ and 
$\pi^+ p$ total cross sections taken from \cite{CERN} are shown, together with the power 
fit of Eq.~(1) (solid line), the fit in ref.~\cite{DL} (dashed curve), and the double 
Reggeon fit of Eq.~(6) with a value $\alpha_{Odd}(0) = 0.9$ (dash-dotted curve).

\begin{figure}[htb]
\centering
\includegraphics[width=.9\hsize]{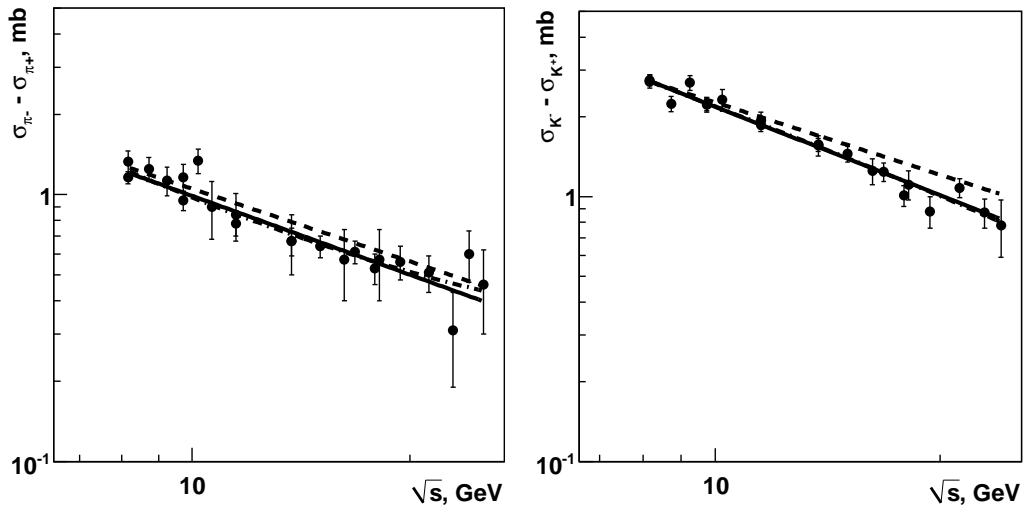}
\vskip -.3cm
\caption{\footnotesize 
The experimental differences in $\pi^- p$ and $\pi^+ p$, left panel (and in $K^- p$ and $K^+ p$, right panel) total cross sections,
together with their fits by Eq.~(1) (solid curves),
by ref.~\cite{DL} (dashed curves), and by Eq.~(6) (dash-dotted curves).}
\end{figure}

Since the Odderon corresponds to a three-gluon exchange, it can not contribute to the 
difference in $\pi^- p$ and $\pi^+ p$ total cross sections, what is consistent with our results,
the values of $\chi^2$/ndf
being practically the same for the solid and 
dash-dotted curves in Fig.~3, while the value of $\sigma_{Odd}$ is compatible with zero. 

Similar results for the differences in $K^- p$ and $K^+ p$ total cross sections are 
shown in the right panel of Fig.~3. Here, the 
double Reggeon fit is again compatible with a zero Odderon contribution.

Needless to say, the presented results do not prove the Odderon existence in $pp$ scattering, 
We can only say that the assumption of the presence of an Odderon contribution is consistent with the experimental data
on total $pp$ and $\bar{p}p$ cross sections. In any case, a more detailed analysis 
is needed, especially concerning the experimental error bars for the differences 
in $pp$ and $\bar{p}p$ cross sections. Thus, we have considered independent experimental values of the $pp$ 
and $\bar{p}p$ cross sections, but the experimental error bars
in their differences would be decreased if both would be measured with the same experimental equipment.

\section{Odderon contribution to the ratio Re/Im parts of elastic $pp$ amplitude}

As one can see from Eqs.~(2) and (3) the Odderon exchange generates a large real part 
of the elastic $pp$ amplitude which is proportional to $\tan({\frac{\pi \alpha_R}2})$.
The singularity at $\alpha_R = 1$ should be compensated by the smallness 
of the corresponding coupling. In the normalization, where $Im A_{hp} = \sigma^{tot}_{hp}$, one
has 
\begin{equation}
Re\hspace{0.05cm}A_{Odd} = \frac12 (\sigma^{tot}_{\bar{p}p} - \sigma^{tot}_{pp})_{Odd}
\cdot\tan\left({\frac{\pi \alpha_R}2}\right)\; ,
\end{equation}
and the additional contribution by the Odderon to the total $\rho = Re A_{pp}/Im A_{pp}$ value
it would be equal to
\begin{equation}
\rho_{Odd} = \frac{Re A_{Odd}}{\sigma^{tot}_{pp}} \;.
\end{equation}

Results in Table 2 and in Fig.~2 show that the possible Odderon contribution to the 
difference in the total $pp$ and $\bar{p}p$ cross sections is of the order of the 
positive signature (mainly Pomeron) contribution at $\sqrt{s} \sim 25$$-$$30$ GeV and 
of about one half of the positive signature contribution at $\sqrt{s} \sim 10$ GeV. 
So, in the case of $\alpha_{Odd}(0) = 0.9$ the value of $Re A_{Odd}$ in Eq.~(8) 
can be $Re A_{Odd}\sim 3$-$4$ mb, what would result in an additional 
$\rho_{Odd} = 0.07$-$0.1$ contribution to the total $Re A_{pp}/Im A_{pp}$ ratio. 
This additional Odderon contribution would be in disagreement with the experimental data
presented in Table~3. 

In fact, the experimental data in refs.~\cite{Vor,Ama} are in good agreement with
the theoretical estimations based on the dispersion relations without Odderon
contribution \cite{Grein}, so the hypothetical Odderon contribution could be as much of the order 
of the experimental error bars. The same situation appears at the CERN-SPS energy~\cite{Aug}.
On the other hand, the experimental points \cite{Faj,Ber} allows some room 
for the presence of the Odderon contribution. It is necessary to keep in mind that the
theoretical predictions also have some ``error bars", e.g. the predictions for UA4
energy presented in \cite{Aug} are between $\rho = 0.12$ \cite{Col} and $\rho = 0.15$ \cite{DL1}.

Let us note that the level of disagreement of the theoretical estimations on $\rho_{Odd}$
with experimental data decreases when decreasing the value of $\alpha_{Odd}$.
\begin{center}
\vskip 5pt
\begin{tabular}{|c||r|r|} \hline
Experiment        & $\rho(s)$\hspace{1.cm} & Theory\hspace{0.75cm} \\  \hline
$\sqrt{s} = 13.7$ GeV \cite{Vor} & $-0.092 \pm 0.014$  & $-0.085$ \cite{Grein}  \\
$\sqrt{s} = 13.7$ GeV \cite{Faj} & $-0.074 \pm 0.018$  & $-0.085$ \cite{Grein}  \\
$\sqrt{s} = 15.3$ GeV \cite{Faj} & $-0.024 \pm 0.014$  & $-0.060$ \cite{Grein}  \\
$\sqrt{s} = 16.8$ GeV \cite{Vor} & $-0.040 \pm 0.014$  & $-0.047$ \cite{Grein}   \\
$\sqrt{s} = 16.8$ GeV \cite{Faj} & $ 0.008 \pm 0.017$  & $-0.047$ \cite{Grein}   \\
$\sqrt{s} = 18.1$ GeV \cite{Faj} & $-0.011 \pm 0.019$  & $-0.04$  \cite{Grein}   \\
$\sqrt{s} = 19.4$ GeV \cite{Faj} & $ 0.019 \pm 0.016$  & $-0.033$ \cite{Grein}   \\
$\sqrt{s} = 21.7$ GeV \cite{Vor} & $-0.041 \pm 0.014$  & $-0.02$  \cite{Grein}  \\
$\sqrt{s} = 23.7$ GeV \cite{Vor} & $-0.028 \pm 0.016$  & $-0.007$  \cite{Grein}   \\
$\sqrt{s} = 30.6$ GeV \cite{Ama} & $ 0.042 \pm 0.011$  &  0.03  \cite{Grein}   \\
$\sqrt{s} = 44.7$ GeV \cite{Ama} & $ 0.062 \pm 0.011$  &  0.062  \cite{Grein}   \\
$\sqrt{s} = 52.9$ GeV \cite{Ama} & $ 0.078 \pm 0.010$  &  0.075  \cite{Grein}   \\
$\sqrt{s} = 62.4$ GeV \cite{Ama} & $ 0.095 \pm 0.011$  &  0.084  \cite{Grein}   \\
$\sqrt{s} = 546$ GeV  \cite{Ber} & $ 0.24 \pm 0.04  $  & 0.10$-$0.15 \cite{Ber}  \\
$\sqrt{s} = 541$ GeV  \cite{Aug} & $ 0.135 \pm 0.015$   & 0.12$-$0.15 \cite{Col,DL1}  \\
\hline
\end{tabular}
\end{center}
Table 3: {\footnotesize Experimental data for the ratio Re/Im parts of elastic $pp$ amplitude
at high energies together with the corresponding theoretical estimations.}

\section{Regge-pole analysis of inclusive particle and antiparticle production in 
the central region}

The inclusive cross section of the production of a secondary $h$ in high energy $pp$
collisions in the central region is determined by the Regge-pole diagrams shown
in Fig.~4~\cite{AKM}. The diagram with only Pomeron exchange (Fig.~4a) is the leading 
one, while the diagrams with one secondary Reggeon $R$ (Figs.~4b and 4c) correspond
to corrections which disappear with the increase of the initial energy.

\begin{figure}[htb]
\centering
\vskip -2.5cm
\includegraphics[width=.45\hsize]{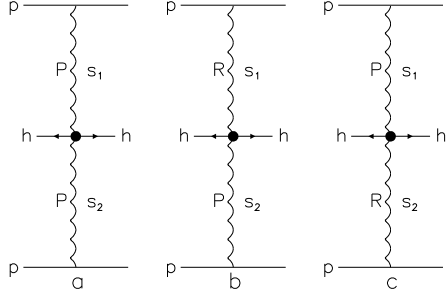}
\vskip -.3cm
\caption{\footnotesize 
Regge-pole diagrams for the inclusive production of a secondary hadron $h$ in the central region.}
\end{figure}

The inclusive production cross section of hadron $h$ with transverse momentum $p_T$
corresponding to the diagram shown in Fig.~4b has the following expression:
\begin{equation}
F(p_T,s_1,s_2,s) = \frac{1}{\pi^2 s} g^{pp}_{R}\cdot g^{pp}_{P}\cdot g^{hh}_{RP}(p_T)\cdot 
\left(\frac{s_1}{s_0}\right)^{\alpha_{R}(0)}\cdot\left(\frac{s_2}{s_0}\right)^{\alpha_{P}(0)} \;,
\end{equation}
where
\begin{eqnarray}
s_1 & = & (p_a + p_h)^2 = m_T\cdot s^{1/2}\cdot e^{-y^*} \\ \nonumber
s_2 & = & (p_b + p_h)^2 = m_T\cdot s^{1/2}\cdot e^{y^*} \;,
\end{eqnarray}
with $s_1\cdot s_2 = m^2_T\cdot s$ \cite{Kar}, and the rapidity $y^*$ defined in the center-of-mass frame. 

The contribution of diagram in Fig.~4c differs from Eq.~(10) in the change of $s_1$ by 
$s_2$ and viceversa, and in the contribution of the diagram in Fig.~4a is obtained from Eq.~(10) by changing
the Reggeon $R$ by Pomeron $P$.

Let us consider the $R$-Reggeon in Fig.~4 as the effective sum of all amplitudes with 
negative signature, so its contribution to the inclusive spectra of secondary protons 
and antiprotons has the opposite sign. In the midrapidity region, i.e. at $y^*=0$, the 
ratios ($\langle m_T \rangle \simeq 1$ GeV) of $p$ and $\bar{p}$ 
yields integrated over $p_T$ can be written as
\begin{equation}
\frac{\bar{p}}p = \frac{1 - r_-(s)}{1 + r_-(s)}  \;,
\end{equation}
where $r_-(s)$ is the ratio of the negative signature ($R$) to the positive signature ($P$) contributions:
\begin{equation}
r_-(s) = c_1\cdot \left(\frac{s}{s_0}\right)^{(\alpha_{R}(0) - \alpha_{P}(0))/2} \;,
\end{equation}
and $c_1$ is a normalization constant. 

The theoretical fit by Eq.~(12) to the experimental data \cite{Gue,Agu,BRA,PHO,PHE,STAR} 
on the ratios of $\bar{p}$ to $p$ production cross sections at $y^*=0$ is presented 
in Fig.~5. Here we have used four experimental points from RHIC, obtained by 
BRAHMS, PHOBOS, PHENIX, and STAR Collaborations, and we present both $\bar{p}/p$ and 
$1 - \bar{p}/p$ as functions of initial energy. The obtained values of the parameters 
$c_1$ and $\alpha_{R}(0) - \alpha_{P}(0)$ are presented in Table 4.

\begin{figure}[htb]
\centering
\includegraphics[width=.9\hsize]{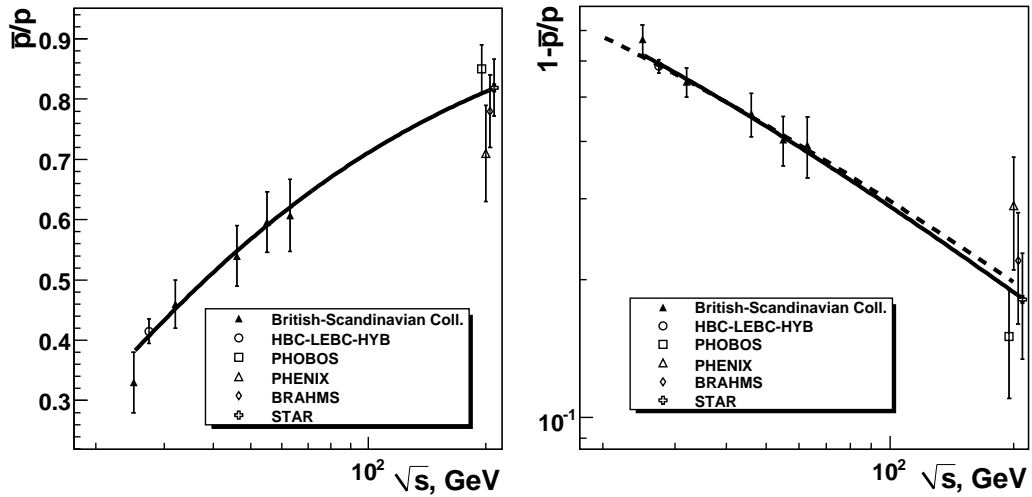}
\caption{\footnotesize 
Ratios of $\bar{p}$ to $p$ production cross sections in high energy $pp$ collisions at $y^*=0$,
together with their fit by Eq.~(12) (solid curves).
Dashed curve shows the result of the fit with only the BRAHMS point at RHIC energy.}
\end{figure}

\vskip 5pt
\begin{center}
\begin{tabular}{|c||r|r|r|} \hline
Parameterization        & $ c_1 $\hspace{1.cm} & $ \alpha_R(0)-\alpha_P(0)$ & $\chi^2$/ndf \\  \hline
$\bar{p}/p$ (Eq.~(13), Fig.~5) & $4.4 \pm 1.1$  & $-0.71 \pm 0.07$  &  4.3/8  \\
$K^-/K^+$ (Eq.~(13), Fig.~6)  & $2.8 \pm 2.6$  & $-0.90 \pm 0.27$  &  2.0/8   \\
$\bar{p}/p$ (Eq.~(14), Fig.~7) & $4.0 \pm 0.7$ & $-0.79 \pm 0.04$ &  15.0/8  \\
$K^-/K^+$ (Eq.~(14), Fig.~7) & $2.3 \pm 0.8$ & $-0.99 \pm 0.12$ &  10.0/7  \\
$\pi^-/\pi^+$ (Eq.~(14), Fig.~7) & $0.44 \pm 0.12$ & $-0.98 \pm 0.11$ &  34.5/7  \\
\hline
\end{tabular}
\end{center}
Table 4: {\footnotesize The Regge-pole fit of the experimental ratios of $\bar{h}p$ and $hp$ total cross
sections by using Eqs.~(12) and (13), and by using Eqs.~(12) and (15).}
\vskip 16pt

The value of difference of $\alpha_R(0)-\alpha_P(0)$ obtained in the fit seems to be too
large for allowing the presence of an Odderon contribution.

The corresponding fit of the experimental data \cite{Gue,Agu,BRA,PHO,PHE,STAR} on the ratios 
of $K^-$ to $K^+$ production cross sections at $y^*=0$ is presented in Fig.~6,
again for $K^-/K^+$ and $1 - K^-/K^+$ as functions of the initial energy. 
The values of the parameters obtained in the fit are also presented in Table 3. The value of 
$\alpha_R(0)-\alpha_P(0)$ obtained in the $K^-/K^+$ is compatible with the value obatined in
the $\bar{p}/p$ fit. 

\begin{figure}[htb]
\centering
\includegraphics[width=.9\hsize]{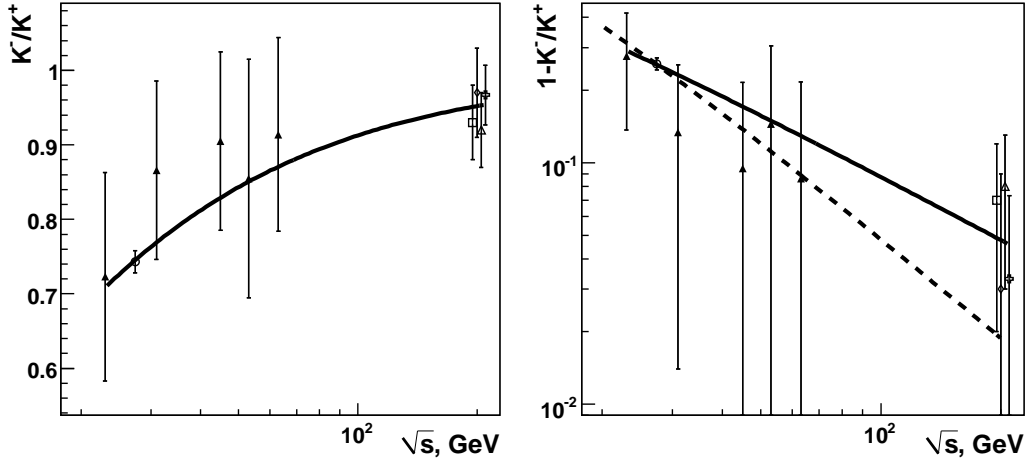}
\vskip -.3cm
\caption{\footnotesize 
Ratios of $K^-$ to $K^+$ production cross sections in high energy $pp$ collisions at 
$y^*=0$, together with their fit by Eq.~(12) (solid curves).
Dashed curve shows the result of the fit with only the BRAHMS point at RHIC energy.}
\end{figure}

It is needed to note that both fits in Figs.~(5) and (6) are in fact normalized to
the experimental point in ref.~\cite{Agu}, since the error bar of this point is several
times smaller than those of the other considered experimental data.

The ratios of $\pi^-$ over $\pi^+$ production cross sections in midrapidity region $y^*=0$ 
differ from unity only at moderate energies where different processes can contribute.
At higher energies, where the applicability of Regge-pole asymptotics seems to be 
reasonable, these ratios are very close to one, so they can not be used in our 
analysis.

Though the experimental points for antiparticle/particle
yield ratios obtained by different Collaborations at RHIC energy $\sqrt{s} = 200$ GeV 
are in reasonable agreement with each other (see Figs.~5 and 6),
the BRAHMS Collaboration results~\cite{BRA1} are of special interest because
they were obtained not only at $y^*=0$,
but also at different values of non-zero rapidity $y^*$, and
they can then provide some additional information. 

Thus we present in the right panels of Figs.~5 and 6 the 
results of the fit to the same experimental data \cite{Gue,Agu} at $\sqrt{s} < 70$ GeV, 
but only considering BRAHMS Collaboration experimental point at RHIC energy (dashed curves).
In the case of the $\bar{p}$ to $p$ ratio the result of this fit is practically the same as with all four 
RHIC points (solid curve in Fig.~5). However in the case of the $K^-$ to $K^+$ ratio the 
fit with only the BRAHMS Collaboration experimental point (dashed curve) significantly 
differs from the solid curve, meaning that the energy dependence of the $K^-$ to $K^+$ experimental ratio is 
very poorly known.

For the case of inclusive production at some rapidity distance $y^* \neq 0$ from the c.m.s.
the quantitiy $r_-(s,y^*)$ in Eq.~(12) takes the form:
\begin{equation}
r_-(s,y^*) = \frac{c_1}2\cdot \left(\frac{s}{s_0}\right)^{(\alpha_{R}(0) - \alpha_{P}(0))/2}\cdot
\left(e^{y^* (\alpha_{R}(0) - \alpha_{P}(0))} + 
e^{-y^* (\alpha_{R}(0) - \alpha_{P}(0))} \right) \;.
\end{equation}

In Fig.~7 we present the fit to the experimental rapidity distribution ratios $\bar{p}/p$ (left panel),
$K^-/K^+$ (right panel), and $\pi^-/\pi^+$ (lower panel) at $\sqrt{s} = 200$ GeV \cite{BRA} by
using Eq.~(14). The values of parameters obtained in this fit are in agreement with those in the
fits of Figs.~5 and~6 (see Table 3), so we can arguably 
claim that in the framework of Regge-pole phenomenology one gets a model independent description of the rapidity dependence of the $\bar{p}/p$ and $K^-/K^+$ ratios by 
using the values of the parameters that were obtained in the description of the energy dependence of these ratios at $y^* = 0$ using Eq.~(12).

\begin{figure}[htb]
\centering
\vskip .4cm
\includegraphics[width=.85\hsize]{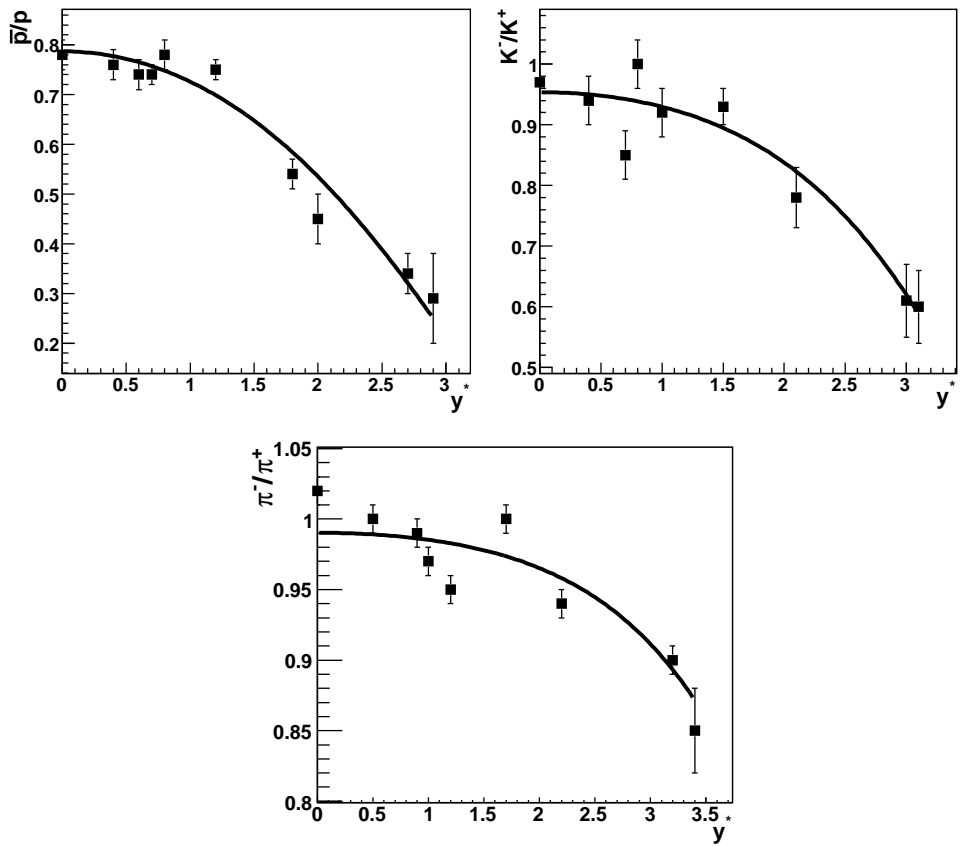}
\vskip -.3cm
\caption{\footnotesize 
Ratios of the inclusive cross sections $\bar{p}$ to $p$ (left panel), $K^-$ to $K^+$ (right panel),
and $\pi^-$ to $\pi^+$ (lower panel) in $pp$ collisions at $\sqrt{s} = 200$ GeV \cite{BRA}
as function of the c.m. rapidity, together with their fit by Eq.~(14) (solid curves).}
\end{figure}

The values of $\alpha_R(0)-\alpha_P(0)$ for the $K^-/K^+$ and $\pi^-/\pi^+$ ratios
are the same and they seem to be larger than the value for the $\bar{p}/p$ ratio. 
This situation is qualitatively similar to that of the differences in the total 
cross sections considered in Section~2. 

However, the fit $\bar{p}/p$ ratios provide values of $\alpha_R(0)-\alpha_P(0)$ 
significantly larger than those one could expect if the Odderon contribution was present.

\section{Inclusive spectra of secondary hadrons \newline in the
Quark-Gluon String Model}

The ratios of inclusive production of different secondaries can also be analyzed
in the framework of the Quark-Gluon String Model (QGSM) \cite{KTM,KaPi,Sh}, which 
allows us to make quantitative predictions at different rapidities including
the target and beam fragmentation regions. In QGSM high energy hadron-nucleon collisions 
are considered as taking place via the exchange of one or several Pomerons, all 
elastic and inelastic processes resulting from cutting through or between Pomerons~\cite{AGK}. 

Each Pomeron corresponds to a cylindrical diagram (see Fig.~8a), and thus, when cutting 
one Pomeron, two showers of secondaries are produced as it is shown in Fig.~8b. The 
inclusive spectrum of a secondary hadron $h$ is then determined by the convolution of 
the diquark, valence quark, and sea quark distributions $u(x,n)$ in the incident 
particles with the fragmentation functions $G^h(z)$ of quarks and diquarks into the 
secondary hadron $h$. These distributions, as well as the fragmentation functions are 
constructed using the Reggeon counting rules \cite{Kai}. Both the diquark and the quark
distribution functions depend on the number $n$ of cut Pomerons in the considered diagram.

\begin{figure}[htb]
\centering
\includegraphics[width=.6\hsize]{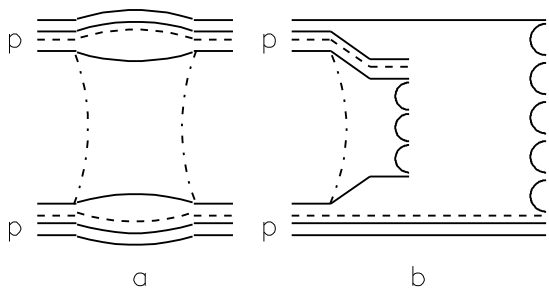}
\vskip -.5cm
\caption{\footnotesize
Cylindrical diagram corresponding to the one--Pomeron exchange contribution to 
elastic $pp$ scattering (a), and the cut of this diagram which determines the 
contribution to the inelastic $pp$ cross section (b). Quarks are shown by 
solid curves and string junction by dashed curves.}
\end{figure}

For a nucleon target, the inclusive rapidity or Feynman-$x$ ($x_F$) spectrum of a 
secondary hadron $h$ has the form~\cite{KTM}:
\begin{equation}
\frac{dn}{dy}\ = \
\frac{x_E}{\sigma_{inel}}\cdot \frac{d\sigma}{dx_F}\ =\ \sum_{n=1}^\infty
w_n\cdot\phi_n^h (x)\ ,
\end{equation}
where the functions $\phi_{n}^{h}(x)$ determine the contribution of
diagrams with $n$ cut Pomerons and $w_n$ is the relative weight of
this diagram. Here we neglect the contribution of diffraction
dissociation processes which is very small in the midrapidity region.

For $pp$ collisions
\begin{eqnarray}
\phi_{pp}^h(x) &=& f_{qq}^{h}(x_+,n)\cdot f_q^h(x_-,n) +
f_q^h(x_+,n)\cdot f_{qq}^h(x_-,n)
\nonumber\\
&&\hspace*{3.5cm}+\ 2(n-1)f_s^h(x_+,n)\cdot f_s^h(x_-,n)\ ,
\\
x_{\pm} &=& \frac12\left[\sqrt{4m_T^2/s+x^2}\ \pm x\right] ,
\end{eqnarray}
where $f_{qq}$, $f_q$, and $f_s$ correspond to the contributions of diquarks, 
valence quarks, and sea quarks, respectively.

These functions are determined by the convolution of the diquark and quark 
distributions with the fragmentation functions, e.g. for the quark one can write:
\begin{equation}
f_q^h(x_+,n)\ =\ \int\limits_{x_+}^1u_q(x_1,n)\cdot G_q^h(x_+/x_1) dx_1\ .
\end{equation}
The diquark and quark distributions, which are normalized to unity, as well 
as the fragmentation functions, are determined by the corresponding Regge intercepts \cite{Kai}.

At very high energies both $x_+$ and $x_-$ are negligibly small in the midrapidity 
region and all fragmentation functions, which are usually written \cite{Kai} as 
$G^h_q(z) = a_h (1-z)^{\beta}$, become constants that are equal for a particle and 
its antiparticle (this would correspond to the limit $r_-(s) \to 0$ in Eq.~(12)):
\begin{equation}
G_q^h(x_+/x_1) = a_h \ . 
\end{equation}
This leads, in agreement with \cite{AKM}, to
\begin{equation}
\frac{dn}{dy}\ = \ g_h \cdot (s/s_0)^{\alpha_P(0) - 1}
\sim a^2_h \cdot (s/s_0)^{\alpha_P(0) - 1} \,,
\end{equation}
that corresponds to the only one-Pomeron exchange diagram in Fig.~4a, the 
only diagram contributing to the inclusive density in the central region (AGK 
theorem \cite{AGK}) at asymptotically high energy. The intercept of the supercritical 
Pomeron $\alpha_P(0) = 1 + \Delta$, $\Delta = 0.139$ \cite{Sh}, is used in the 
QGSM numerical calculations. 

\section{The baryon as a 3q+SJ system}

In the string models, baryons are considered as configurations consisting of 
three connected strings (related to three valence quarks) called  string 
junction  (SJ) \cite{Artru,IOT,RV,Khar}. The colour part of a baryon 
wave function reads as follows~\cite{Artru,RV} (see Fig.~9):

\begin{figure}[htb]
\centering
\vskip -1.5cm
\includegraphics[width=.5\hsize]{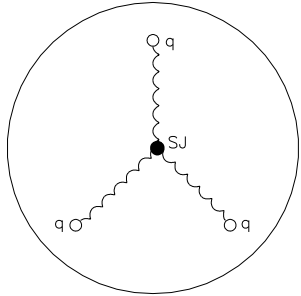}
\vskip -.8cm
\caption{\footnotesize
The composite structure of a baryon in string models. Quarks are shown by open 
points.}
\end{figure}
\begin{eqnarray}
&&B\ =\ \psi_i(x_1)\cdot\psi_j(x_2)\cdot\psi_k(x_3)\cdot J^{ijk}(x_1, x_2, x_3, x) 
\,,
\\
&& J^{ijk}(x_1, x_2, x_3, x) =\ \Phi^i_{i'}(x_1,x)\cdot\Phi_{j'}^j(x_2,x)\cdot
\Phi^k_{k'}(x_3,x)\cdot\epsilon^{i'j'k'} \,,
\\
&& \Phi_i^{i'}(x_1,x) = \left[ T\cdot\exp \left(g\cdot\int\limits_{P(x_1,x)}
A_{\mu}(z) dz^{\mu}\right) \right]_i^{i'} \,,
\end{eqnarray}
where $x_1, x_2, x_3$, and $x$ are the coordinates of valence quarks and SJ, 
respectively, and $P(x_1,x)$ represents a path from $x_1$ to $x$ which looks 
like an open string with ends at $x_1$ and $x$. Such a baryon structure is supported
by lattice calculations \cite{latt}. 

This picture leads to some general phenomenological predictions. In
particular, it opens room for exotic states, such as the multiquark bound 
states, 4-quark mesons, and pentaquarks \cite{RV,DPP1,RSh}. In the case of 
inclusive reactions the baryon number transfer to large rapidity distances in 
hadron-nucleon and hadron-nucleus reactions can be explained 
\cite{ACKS,BS,AMS,Olga,SJ3,AMS1} by SJ diffusion.

The production of a baryon-antibaryon pair in the central region usually occurs
via $SJ$-$\overline{SJ}$ (according to Eqs.~(21) and (22) SJ has upper color indices,
whereas antiSJ ($\overline{SJ}$) has lower indices) pair production which then combines 
with sea quarks and sea antiquarks into, respectively, $B\bar{B}$ pair~\cite{RV,VGW},
as it is shown in Fig.~(10). In the case of $pp$ collisions the existence of two SJ
in the initial state and their diffusion in rapidity space lead to significant differences
in the yields of baryons and antibaryons in the midrapidity region even at rather high
energies~\cite{ACKS,AMS}.

\begin{figure}[htb]
\centering
\vskip -3.5cm
\hspace{4.cm}
\includegraphics[width=.6\hsize]{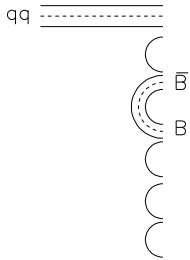}
\vskip -1.8cm
\caption{\footnotesize Diagram corresponding to the diquark fragmentation function for the
production of a central $\bar{B}B$ pair. Quarks are shown by solid curves and SJ by
dashed curves.}
\end{figure}
 
The quantitative theoretical description of the baryon number transfer via SJ 
mechanism was suggested in the 90's and used to predict~\cite{KP1} the
$p/\bar{p}$ asymmetry at HERA energies then experimentally observed~\cite{H1}.
Also in ref.~\cite{Bopp} was noted that the $p/\bar{p}$ asymmetry measured at HERA
can be obtained by simple extrapolation of ISR data.
The quantitative description of the baryon number transfer due to SJ diffusion in 
rapidity space was obtained in~\cite{ACKS} and following
papers~\cite{ACKS,BS,AMS,Olga,SJ3,AMS1,MRS}. 

In the QGSM the differences in the spectra of secondary baryons and antibaryons
produced in the central region appear for processes which present SJ diffusion
in rapidity space. These differences only vanish rather slowly when the energy
increases. 

To obtain the net baryon charge, and
according to ref.~\cite{ACKS}, we consider three different possibilities.
The first one is the fragmentation of the diquark 
giving rise to a leading baryon (Fig.~11a). A second possibility is to produce 
a leading meson in the first break-up of the string and a baryon in a 
subsequent break-up~\cite{Kai,22r} (Fig.~11b). In these two first cases the baryon 
number transfer is possible only for short distances in rapidity. In the third 
case, shown in Fig.~11c, both initial valence quarks recombine with sea
antiquarks into mesons $M$ while a secondary baryon is formed by the SJ together 
with three sea quarks.

\begin{figure}[htb]
\centering
\includegraphics[width=.55\hsize]{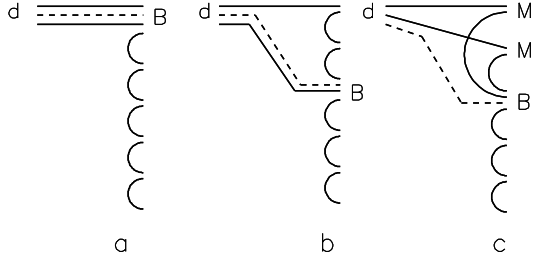}
\vskip -0.1cm
\caption{\footnotesize
QGSM diagrams describing secondary baryon $B$ production by diquark $d$: 
initial SJ together with two valence quarks and one sea quark (a), initial SJ 
together with one valence quark and two sea quarks (b), and initial SJ 
together with three sea quarks (c).}
\end{figure}

The fragmentation functions for the secondary baryon $B$ 
production corresponding to the three processes shown in Fig.~11 can be written
as follows (see~\cite{ACKS} for more details):
\begin{eqnarray}
G^B_{qq}(z) &=& a_N\cdot v_{qq} \cdot z^{2.5} \;,
\\
G^B_{qs}(z) &=& a_N\cdot v_{qs} \cdot z^2\cdot (1-z) \;,
\\
G^B_{ss}(z) &=& a_N\cdot\varepsilon\cdot v_{ss} \cdot z^{1 - \alpha_{SJ}}\cdot (1-z)^2 \;,
\end{eqnarray}
for Figs.~11a, 11b, and 11c, respectively, and where $a_N$ 
is the normalization parameter, and $v_{qq}$, $v_{qs}$, $v_{ss}$ are the 
relative probabilities for different baryons production that can be found by
simple quark combinatorics \cite{AnSh,CS}. 

The fraction $z$ of the incident baryon energy carried by the secondary baryon decreases 
from Fig.~11a to Fig.~11c, whereas the mean rapidity gap between the incident and 
secondary baryon increases. The first two processes can not contribute to the inclusive 
spectra in the central region, but the third contribution is essential if the value of 
the intercept of the SJ exchange Regge-trajectory, $\alpha_{SJ}$, is large enough.
At this point it is important to stress that since the quantum number content of the $SJ$
exchange matches that of the possible Odderon exchange, if the value of the $SJ$
Regge-trajectory intercept, $\alpha_{SJ}$, would turn out to be large and it would 
coincide with the value of the Odderon Regge-trajectory, $\alpha_{SJ}\simeq 0.9$, 
then the $SJ$ could be identified to the Odderon, or to one baryonic Odderon component.

Let's finally note that the process shown in Fig.~11c can be very naturally realized 
in the quark combinatorial approach~\cite{AnSh} through the specific probabilities of
a valence quark recombination (fusion) with sea quarks and antiquarks, the
value of $\alpha_{SJ}$ depending on these specific probabilities.

The contribution of the graph in Fig.~11c has in QGSM a coefficient $\varepsilon$ 
which determines the small probability for such a baryon number transfer to occur.

At high energies the SJ contribution from Eq.~(26) to the inclusive cross section of
secondary baryon production at large rapidity distance $\Delta y$ from the incident
nucleon can be estimated as
\begin{equation}
\frac{1}{\sigma}\cdot\frac{d\sigma^B}{dy} \sim a_B\cdot\varepsilon\cdot
e^{(1 - \alpha_{SJ}) \Delta y} \;,
\end{equation}
where $a_B = a_N\cdot v^B_{ss}$. The baryon charge transferred to large rapidity
distances can be determined by integration of Eq.~(27), so being of the order of
\begin{equation}
\langle n_B \rangle \sim \frac{a_B\cdot\varepsilon}{(1 - \alpha_{SJ})} \;.
\end{equation}
It is then clear that the value $\alpha_{SJ} \geq 1$ should be excluded due to the
violation of baryon-number conservation at asymptotically high energies.

\section{Comparison of the QGSM predictions to the experimental data}

With the value $\alpha_{SJ}=0.5$ used to obtain the first
QGSM predictions~\cite{ACKS} different values of $\varepsilon$ in Eq.~(26) were needed for the correct 
description of the experimental data at moderate and high energies. A better solution 
was found in ref.~\cite{BS}, where it was shown that all experimental data can be described with 
the value $\alpha_{SJ}=0.9$ and only one value of $\varepsilon$.
This large value of $\alpha_{SJ}$ allows to 
describe the preliminary experimental data of H1 Collaboration~\cite{H1} on asymmetry 
of $p$ and $\bar{p}$ production in $\gamma p$ interactions at HERA with a rather small
change in the description of the data at moderate energies. A similar analysis presented
in ref.~\cite{Olga} for midrapidity asymmetries of
$\bar{\Lambda}/\Lambda$ produced in $pp$, $pA$, $\pi p$, and $ep$ interactions
also shows that the value $\alpha_{SJ}=0.9$ is slightly favoured,
mainly due to the H1 Collaboration point~\cite {H1a}.

Here we compare the results of QGSM predictions with all available experimental data 
on the $\bar{p}/p$ ratios presented in Fig.~5. To obtain these predictions we use the 
values of the probabilities $w_n$ in Eq.~(15) that are calculated in the frame of 
Reggeon theory \cite{KTM}, and the values of the normalization constants $a_{\pi}$ 
(pion production), $a_K$ (kaon production), $a_{\bar{N}}$ ($B\bar{B}$ pair production), 
and $a_N$ (baryon production due to SJ diffusion) that were determined~\cite{KTM,KaPi,Sh}
from the experimental data at fixed target energies. 

To compare the QGSM results obtained with different values of $\alpha_{SJ}$ in Eq.~(26)
all curves should be normalized at the same arbitrary point. This can be done in two
different ways, either to use the experimental points at the highest energy, or
to use the experimental point with the smallest error bar.

In the first case one takes into account that Reggeon theory is an asymptotic theory,
and so numerically small (say of the order of 10\%) disagreements with the experimental
data at fixed target energies are not very important. This first possibility was used
in~\cite{BS,AMS1} to obtain a
rather good description of high energy data (HERA, RHIC, and ISR), and a reasonable
agreement with the data at fixed target energies. However, the value of $\chi^2$ calculated
here from the data at fixed target energies is not that good.

As example we present in Table 5 the calculated yields of different secondaries produced
in $pp$ collisions at RHIC energy $\sqrt{s} = 200$ GeV~\cite{AMS1}.

\begin{center}
\begin{tabular}{|c||r|r|r|} \hline
Particle & \multicolumn{3}{c|}{RHIC ($\sqrt{s} = 200$ GeV)}  \\ \cline{2-4}
& $\varepsilon = 0$ & $\varepsilon = 0.024$ & STAR Collaboration \cite{abe} \\  \hline
$\pi^+$         & 1.27    &        &                    \\
$\pi^-$         & 1.25    &        &                    \\
$K^+$           & 0.13    &        & $0.14 \pm 0.01$    \\
$K^-$           & 0.12    &        & $0.14 \pm 0.01$    \\
$p$             & 0.0755  & 0.0861 &                    \\
$\overline{p}$  & 0.0707  &        &                    \\
$\Lambda$       & 0.0328  & 0.0381 & $0.0385 \pm 0.0035$  \\
$\overline{\Lambda}$ & 0.0304  &   & $0.0351 \pm 0.0032$  \\
$\Xi^-$      & 0.00306  & 0.00359 & $0.0026 \pm 0.0009$   \\
$\overline{\Xi}^+$ & 0.00298 &     & $0.0029 \pm 0.001$   \\
$\Omega^-$       & 0.00020  & 0.00025 & *                 \\
$\overline{\Omega}^+$ & 0.00020  &    & *                 \\
\hline
\end{tabular}
\end{center}
\vspace{-0.2cm}

\noindent
\hspace{2.5cm}$^* dn/dy(\Omega^- + \overline{\Omega}^+) = 0.00034 \pm 0.00019$

\noindent
Table 5: {\footnotesize The QGSM results ($\alpha_{SJ} = 0.9$) for midrapidity
yields $dn/dy$ ($\vert y \vert < 0.5$) integrated over $p_T$
for different secondaries at RHIC energies. The results for $\varepsilon = 0.024$ are
presented only when different from the case $\varepsilon = 0$.}

The second way was used in~\cite{MRS}, where it was possible to describe all the data
with good $\chi^2$, except for the $\bar{p}p$ and $\bar{\Lambda}\Lambda$ asymmetries
at HERA (see discussion below). In this case the experimental value of $\bar{p}/p$
ratio at $\sqrt{s}$ = 27.4 GeV~\cite{Agu} was chosen as normalization. 
To do so it was necessary to slightly change the fragmentation function of 
$uu$ and $ud$ diquarks into secondary antiproton, which now has the form:
\begin{equation}
G^{\bar{p}}_{uu}(z) = G^{\bar{p}}_{ud}(z) =
a_{\bar{N}}\cdot (1 - z)^{\lambda - \alpha_R + 4(1-\alpha_B)}\cdot (1 + 3z) \;,
\end{equation}
with a smaller value of $a_{\bar{N}}$, $a_{\bar{N}} = 0.13$, and an additional factor 
$(1 + 3z)$ with respect to the expression in ref.~\cite{ACKS}.
For all other quark distributions and fragmentation functions the same
expressions as in ref.~\cite{ACKS} have been taken. The quality of the description of
the $\bar{p}$ inclusive spectra with fragmentation function of Eq.~(29) is shown to
be even better than in previous papers (see Fig.~12). Here, however, the calculated
inclusive density of antiprotons at RHIC energy decreases in comparison with the
value presented in Table 5.

\begin{figure}[htb]
\centering
\includegraphics[width=.48\hsize]{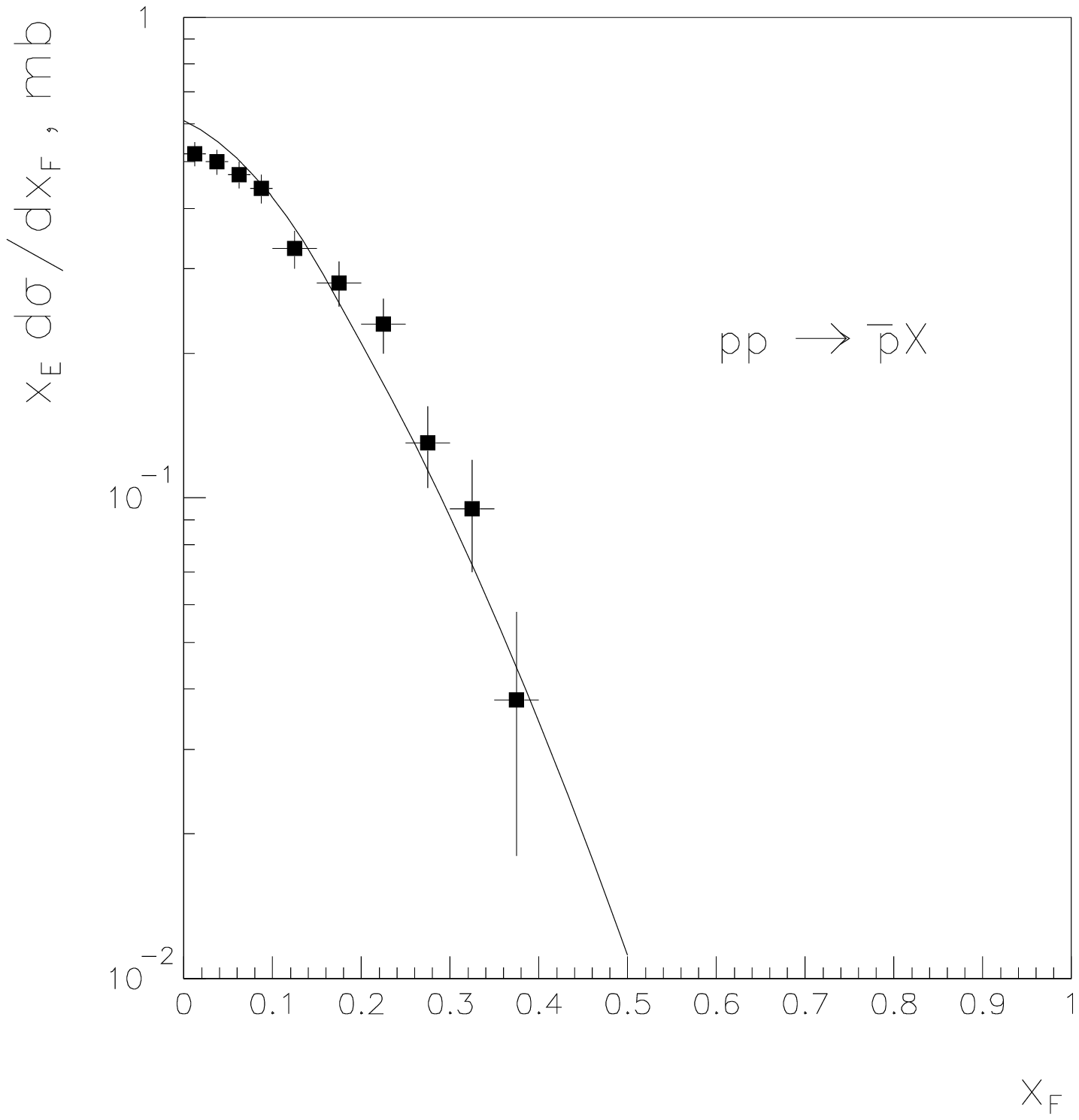}
\includegraphics[width=.48\hsize]{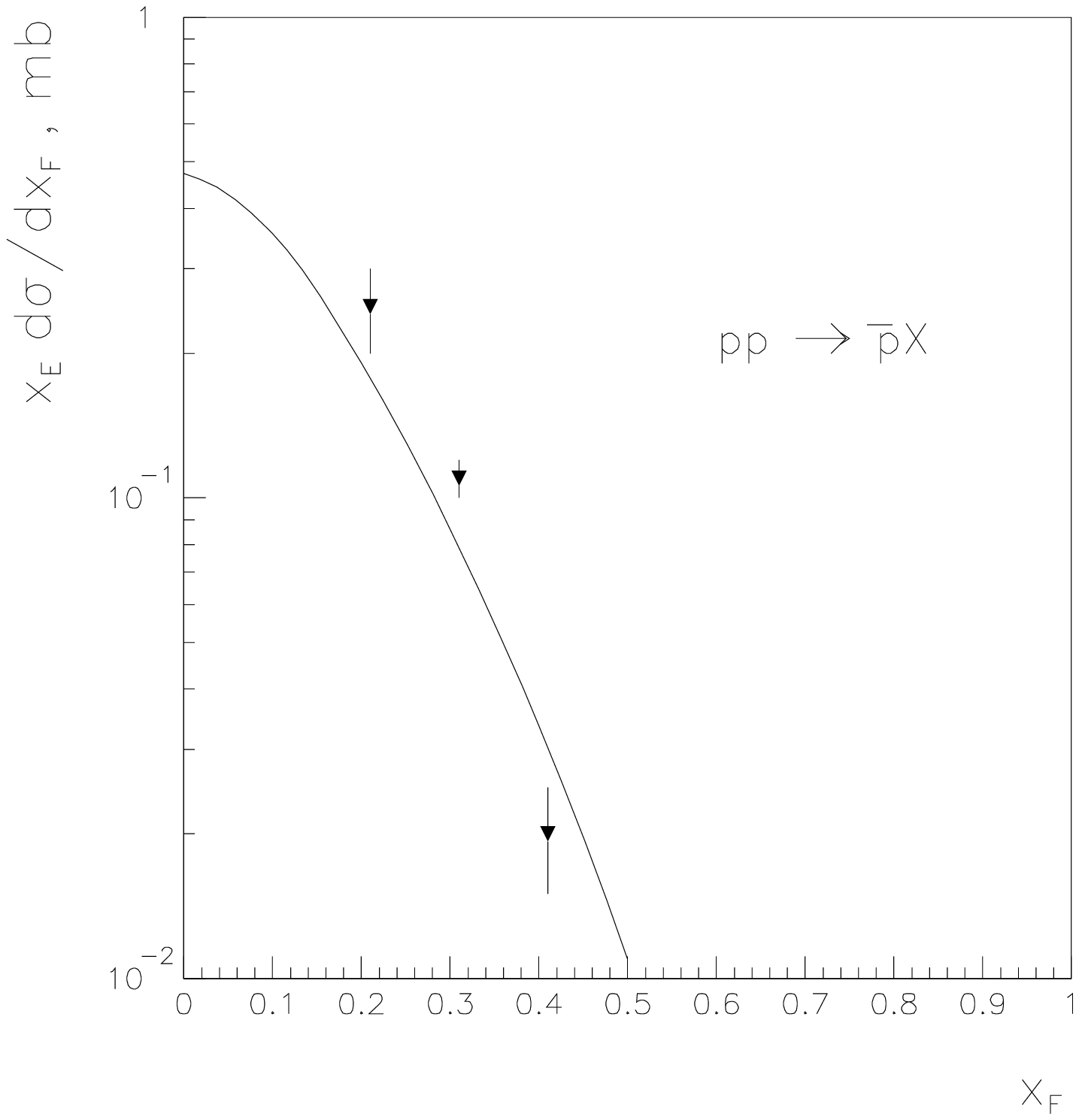}
\vskip -.3cm
\caption{\footnotesize 
The experimental spectra of $\bar{p}$ produced in $pp$ collisions at 400 GeV/c \cite{Agu}
(left panel) and 175 GeV/c \cite{Bren} (right panel),
together with their QGSM description.}
\end{figure}

The ratio of $p$ to $\bar{p}$ yields at $y^*=0$ calculated with the QGSM and with the
fragmentation function in Eq.~(29) is shown in the 
left panel of Fig.~13. The results with $\alpha_{SJ} = 0.9$ and $\varepsilon = 0.024$, 
$\alpha_{SJ} = 0.6$ and $\varepsilon = 0.057$, and $\alpha_{SJ} = 0.5$ and $\varepsilon = 0.0757$
are presented by dashed ($\chi^2$/ndf=21.7/10), dotted ($\chi^2$/ndf=12.2/10), 
and dash-dotted ($\chi^2$/ndf=11.1/10) curves, respectively.
Thus the most probable value of $\alpha_{SJ}$ from the point of view 
of $\chi^2$ analyses is $\alpha_{SJ} = 0.5 \pm 0.1$.

\begin{figure}[htb]
\centering
\includegraphics[width=.9\hsize]{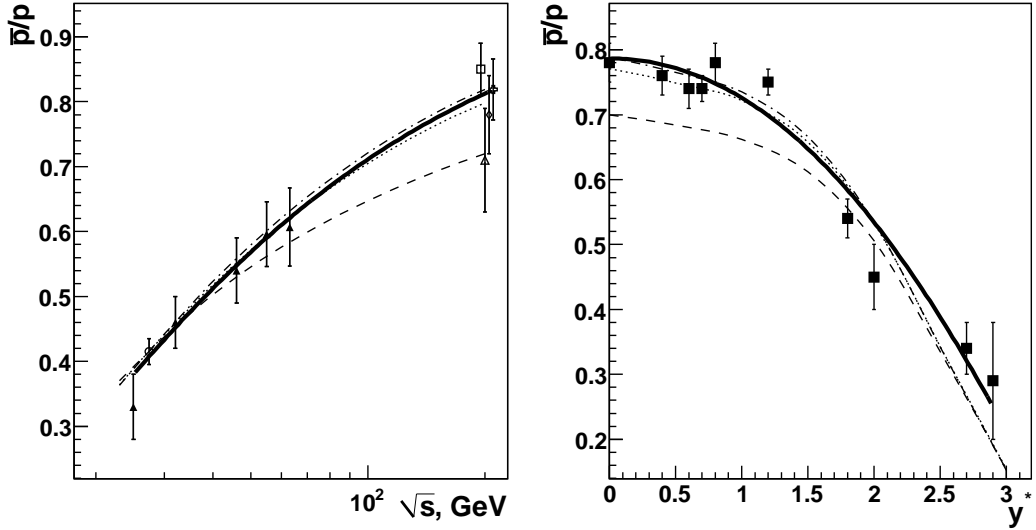}
\caption{\footnotesize 
The experimental ratios of $\bar{p}$ to $p$ production cross sections in high energies 
$pp$ collisions at $y^*=0$ \cite{Gue,Agu,BRA,PHO,PHE,STAR} (left panel) and as the 
functions of rapidity at $\sqrt{s} = 200$ GeV \cite{BRA} (right panel), together with 
their fits by Eqs.~(12), (13), and (14) (solid curves), and by the QGSM description 
(dashed, dotted, and dash-dotted curves).}
\end{figure}

The calculated ratios of $\bar{p}$ to $p$ yields as function of rapidity are shown in the right panel of Fig.~12.
In accordance with the experimental conditions~\cite{BRA} we use 
here the value $\langle p_T \rangle$ = 0.9 GeV/c both for secondary $p$ and $\bar{p}$. 
We also present here the calculations with $\alpha_{SJ} = 0.9$ and $\varepsilon = 0.024$, $\alpha_{SJ} = 0.6$
and $\varepsilon = 0.057$, and $\alpha_{SJ} = 0.5$ and $\varepsilon = 0.0757$ by dashed ($\chi^2$/ndf=72.8/10),
dotted ($\chi^2$/ndf=18.6/10), and dash-dotted 
($\chi^2$/ndf=17.0/10) curves, the most probable value of $\alpha_{SJ}$ being again 
$0.5 \pm 0.1$.

\section{Predictions for LHC}

We present in Tables 6 and 7 our predictions for antibaryon/baryon ratios in
midrapidity region at energies $\sqrt{s} = 900$ GeV and $\sqrt{s} = 14$ TeV by
considering three different scenarios for the SJ contribution: not SJ contribution
($\varepsilon = 0$), SJ contribution with $\alpha_{SJ} = 0.5$, and SJ contribution
with $\alpha_{SJ} = 0.9$.

The first remarkable feature of our prediction is that practically equal $\bar{B}/B$ ratios
are predicted for baryons with different strangeness content, the small differences presented
in Tables 6 and 7 being inside the of the accuracy of our calculations.

At $\sqrt{s} = 900$ GeV, probably the lowest energy used for testing at LHC as intermediate energy
between RHIC and LHC energies, we expect the values of $\bar{B}/B$ ratios to be about 0.96 in the case
of $\alpha_{SJ} = 0.5$ (which corresponds to the well known $\omega$-reggeon exchage at lower energies),
and of about 0.90 in the case of $\alpha_{SJ} = 0.9$ (Odderon contribution).

\begin{center}
\vskip 5pt
\begin{tabular}{|c||r|r|r|} \hline
Ratio & \multicolumn{3}{c|}{$\sqrt{s} = 900$ GeV}  \\ \cline{2-4}
& $\varepsilon = 0$ & $\alpha_{SJ} = 0.5$ & $\alpha_{SJ} = 0.9$  \\   \hline
$\overline{p}/p$ & 0.981 & 0.955 & 0.892  \\
$\overline{\Lambda}/\Lambda$ & 0.976 & 0.949 & 0.887  \\
$\overline{\Xi}^+/\Xi^-$ & 0.992 & 0.965 & 0.909  \\
$\overline{\Omega}^+/\Omega^-$ & 1.0 & 0.965 & 0.907  \\
\hline
\end{tabular}
\end{center}
Table 6: {\footnotesize The QGSM predictions for antibaryon/baryon yields ratios in $pp$ collisions
in midrapidity region ($\vert y \vert < 0.5$) at $\sqrt{s} = 900$ GeV. Three different scenarios
for the SJ contribution are considered: not SJ contribution ($\varepsilon = 0$), SJ contribution
with $\alpha_{SJ} = 0.5$ ($\omega$-reggeon exchange), and SJ contribution with $\alpha_{SJ} = 0.9$
(Odderon exchange).}

For $\sqrt{s} = 14$ TeV the $\bar{B}/B$ ratios are predicted to be larger
than 0.99 for $\alpha_{SJ} = 0.5$ and of about 0.96 for $\alpha_{SJ} = 0.9$,
so an experimental accuracy of the order of $\sim 1$ \% in the measurement of these ratios will be
needed to discriminate between the two values of $\alpha_{SJ}$.

\begin{center}
\vskip 5pt
\begin{tabular}{|c||r|r|r|} \hline
Ratio & \multicolumn{3}{c|}{$\sqrt{s} = 14$ TeV} \\ \cline{2-4}
&  $\varepsilon = 0$ & $\alpha_{SJ} = 0.5$ & $\alpha_{SJ} = 0.9$ \\   \hline
$\overline{p}/p$ & 0.998 & 0.994 & 0.957  \\
$\overline{\Lambda}/\Lambda$  & 0.992 & 0.993 & 0.957 \\
$\overline{\Xi}^+/\Xi^-$  & 0.999 & 0.995 & 0.966  \\
$\overline{\Omega}^+/\Omega^-$ & 1.0 & 0.995 & 0.967  \\
\hline
\end{tabular}
\end{center}
Table 7: {\footnotesize The QGSM predictions for antibaryon/baryon yields ratios 
in $pp$ collisions in midrapidity region ($\vert y \vert < 0.5$) at $\sqrt{s} = 14$~TeV.
Three different scenarios for the SJ contribution are considered: not SJ contribution
($\varepsilon = 0$), SJ contribution with $\alpha_{SJ} = 0.5$ ($\omega$-reggeon exchange),
and SJ contribution with $\alpha_{SJ} = 0.9$ (Odderon exchange).}

\vskip 15pt

Our predictions for the $\bar{B}/B$ ratios at intermediate LHC energies
$\sqrt{s} = 5.5$ TeV and $\sqrt{s} = 10$ TeV, calculated with Odderon contribution
$\alpha_{SJ} = 0.9$, are presented in Table 8.

\newpage

\begin{center}
\begin{tabular}{|c||r|r|} \hline
Ratio & $\sqrt{s} = 5.5$ TeV & $\sqrt{s} = 10$ TeV  \\  \hline
$\overline{p}/p$ & 0.942 & 0.953  \\
$\overline{\Lambda}/\Lambda$  & 0.941 & 0.953 \\
$\overline{\Xi}^+/\Xi^-$  & 0.954 & 0.962  \\
$\overline{\Omega}^+/\Omega^-$ & 0.954 & 0.962  \\
\hline
\end{tabular}
\end{center}
Table 8: {\footnotesize The QGSM predictions for antibaryon/baryon yields ratios
in $pp$ collisions in midrapidity region ($\vert y \vert < 0.5$) for energies
$\sqrt{s} = 5.5$ TeV and $\sqrt{s} = 10$ TeV, obtained by considering one
Odderon contribution with $\alpha_{SJ} = 0.9$.}

\vskip 15pt

The midrapidity densities of produced secondaries are more model dependent than the
antiparticle/particle ratios. We present in Table 9 the QGSM predictions at LHC
energies $\sqrt{s} = 5.5$ TeV, $\sqrt{s} = 10$ TeV, and $\sqrt{s} = 14$ TeV,
obtained by considering one Odderon contribution with $\alpha_{SJ} = 0.9$ and
$\varepsilon = 0.024$. Baryon densities estimated without any Odderon contribution
are smaller, and they practically coincide with the antibaryon densities at the same energy.

\begin{center}
\begin{tabular}{|c||r|r|r|} \hline
Particle &  $\sqrt{s} = 5.5$ TeV & $\sqrt{s} = 10$ TeV
& LHC ($\sqrt{s} = 14$ TeV) \\   \hline
$\pi^+$         & 2.23   & 2.43    & 2.54  \\
$\pi^-$         & 2.23   & 2.43    & 2.54   \\
$K^+$           & 0.22   & 0.24    & 0.25   \\
$K^-$           & 0.22   & 0.24    & 0.25   \\
$p$             & 0.161  & 0.176   & 0.185  \\
$\overline{p}$  & 0.152  & 0.168   & 0.177  \\
$\Lambda$       & 0.0784 & 0.0868  & 0.0914 \\
$\overline{\Lambda}$  & 0.0738   & 0.0827   & 0.0875   \\
$\Xi^-$               & 0.0094   & 0.0106   & 0.0113   \\
$\overline{\Xi^+}$    & 0.0089   & 0.0102   & 0.0109   \\
$\Omega^-$            & 0.00076  & 0.00088  & 0.00094  \\
$\overline{\Omega^+}$ & 0.00073  & 0.000845 & 0.000912  \\
\hline
\end{tabular}
\end{center}
Table 9: {\footnotesize The QGSM results for midrapidity yields $dn/dy$ ($\vert y \vert < 0.5$)
of different secondaries at LHC energies  $\sqrt{s} = 5.5$ TeV, $\sqrt{s} = 10$ TeV,
and $\sqrt{s} = 14$ TeV, obtained by considering one Odderon contribution with $\alpha_{SJ} = 0.9$.}

\section{Discussion and conclusions}

From the general point of view, our approach, in which the Odderon effects in
soft interactions are generated by $t$-channel SJ exchange, coincides with those in
Refs.~\cite{BLV,IOT,RV,Khar,KP1,KP2}. Really, the two-Pomeron diagram in Fig.~4a
leads by definition to equal yields of baryons and antibaryons in the central region,
because the Pomeron has vacuum quantum numbers. All the difference in the production
of  baryons and antibaryons has to come from the diagrams Figs.~4b and/or 4c, where
the Reggeon $R$ has negative C-parity and negative signature, so it contributes to the
baryon and antibaryon production with opposite signs. In string models, such contribution
is identified with SJ diffusion in rapidity space in the inelastic amplitude of particle
production, and, due to unitarity, these squared inelastic amplitudes result in the diagrams
of Figs.~4b, or 4c. If $\alpha_{SJ} \leq 0.6$ the $R$-reggeon with negative signature
should be identified to the $\omega$-Reggeon, while in the case of $\alpha_{SJ} \geq 0.7-0.8$
it should be correspond to the Odderon. From this point of view, the hard Odderon represents
the triple-gluon exchange~\cite{BLV,KP1,KP2}, whereas the soft Odderon consists of
a SJ-antiSJ exchange.

Some differences between our the numerical results and those presented 
in~\cite{BLV,IOT,RV,Khar,KP1,KP2} can be connected to the processes (diagrams) which are 
considered in the calculations by each approach, and to the values of the 
corresponding model parameters. Also, the values of $\alpha_{SJ}(0)$ extracted from 
the experimental data depend on the energy region and on the experimental
points included into the analysis.

We neglect the possibility of interactions between Pomerons (the so-called enhancement
diagrams) in our calculations. Such interactions are already very important in the 
cases of heavy ion~\cite{CKT} and nucleon-nucleus~\cite{MPS} collisions at RHIC 
energies, and their contribution should increase with energy. However, we estimate 
that the inclusive density of secondaries produced in $pp$ collisions at LHC energies 
is not large enough to affected by these effects.

In the case of heavy ion collisions, the effects of SJ diffusion ($\bar{B}B$ asymmetry and others)
could be increased by the contribution to the inclusive spectrum of a secondary baryon of diagrams
in which several incident nucleons interact with the target to the inclusive spectrum of a
secondary baryon. The detailed study of this contribution will be a requirement of further analysis.

The experimental data on the differences in particle and antiparticle total cross 
sections with a proton target have been considered in Section 2. As discussed there, 
a possible Odderon contribution can be present in this case of $\bar{p}p$ and $pp$ total cross sections,
while such an Odderon contribution shoud turn out to be zero for meson-proton total cross sections.
With the Odderon corresponding to a reggeized three-gluon exchange in $t$-channel this last
feature naturally appears.

However one has to note that the possibility of an Odderon exchange contribution 
in $\bar{p}p$ and $pp$ is supported by the data of $\bar{p}/p$ scattering obtained at
ISR energies ($\sqrt{s} \geq  30$ GeV), where experimental data (at so high energies)
do not exist for $K^{\pm}p$ and $\pi^{\pm}p$ exist.

On the other hand, the main part of experimental data for the ratios of real/imaginary 
parts of elastic $pp$ amplitude, including the ISR data, are in agreement, either with the absence 
of any Odderon contribution, or with the presence of a very small Odderon contribution,
if the value of $\alpha_{Odd}$ is close to one. The exceptions 
to this fact are the FNAL data \cite{Faj} and the oldest CERN-SPS experimental point~\cite{Ber}
that allows some room for the Odderon contribution to be present.

Thus it seems that the ISR data on the differences of particle and antiparticle total
interaction cross sections and the data on the ratios of real/imaginary parts of 
elastic $pp$ amplitude are not completely consistent with each other. 

In the case of the inclusive production of particles and antiparticles in central 
(midrapidity) region in $pp$ collisions we could not see any contribution by the Odderon. 
All experimental data starting from those at fixed target energies are consistent with
a value $\alpha_R(0) \simeq 0.5$, a little larger than the conventional value of
$\alpha_{\omega}(0) \simeq 0.4$, but still too small for the Odderon contribution to be
there. On top of that, even at RHIC energies the available energy for the possible
Odderon exchange is $\sqrt{s}\simeq 15$$-$$20$ GeV, what is perhaps too small, since
we did not saw any Odderon contribution at such energies in the case of the differences
of particle and antiparticle total interaction cross sections. Actually, the only
evidence for the Odderon exchange with $\alpha_{Odd}(0) \simeq 0.9$ in inclusive
reactions are two experimental points for $\bar{B}B$ production asymmetry by the
H1 Collaboration~\cite{H1,H1a}. The first point \cite{H1} (for $\bar{p}/p$ ratio)
is until now not published, and the second one~\cite{H1a}
(for $\bar{\Lambda}/\Lambda$ ratio) shows a very large error bar, but, on the 
other hand, only for these two points the kinematics would allow the energy of the 
Odderon exchange to be large enough, $\sqrt{s} \simeq 100$ GeV.

One has to expect that the LHC data will make the situation more clear. The QGSM 
predictions for the deviation of $\bar{B}/B$ ratios from unity due to SJ 
contribution with $\alpha_{SJ}(0)\simeq 0.9$ have been already published~\cite{AMS1}
(see also Tables 6, 7, 8, and 9 in section~8), and they allow deviations from unity
on the level of $\sim 4$\% at $\sqrt{s} = 14$ TeV,
while for smaller values of $\alpha_{SJ}(0)$ these ratios should be close to one.

\newpage

{\bf Acknowledgements}

We are grateful to A.B. Kaidalov for the idea of providing this analysis and
discussions, and to Ya.I. Azimov and M.G. Ryskin for useful discussions and comments.
We also thank Y. Foka and P. Christakoglou for sharing with us their insight on the
experimental requirements and conditions at LHC. 

This paper was supported by Ministerio de Educaci\'on y Ciencia of Spain under the 
Spanish Consolider-Ingenio 2010 Programme CPAN (CSD2007-00042) and project FPA 2005--01963, 
by Xunta de Galicia and, in part, by grants RFBR-07-02-00023 and RSGSS-1124.2003.2.


\end{document}